\documentstyle[12pt]{article} 
\input psfig.sty

\def\Title#1{\begin{center} {\Large #1 } \end{center}}
\def\Author#1{\begin{center}{ \sc #1} \end{center}}
\def\Address#1{\begin{center}{ \it #1} \end{center}}

\def\doeack{\footnote{Work supported by the Department of Energy,
                     contract DE--AC03--76SF00515.}}
\def\SLAC{Stanford Linear Accelerator Center\\
    Stanford University, Stanford, California 94309 USA}

\newenvironment{Abstract}{\begin{quotation} \begin{center}
                       ABSTRACT
     \end{center}\bigskip  }{\end{quotation}}
\def\beq{\begin{equation}}
\def\eeq#1{\label{#1}\end{equation}}
\def\eeqn{\end{equation}}
\def\beqa{\begin{eqnarray}}
\def\eeqa#1{\label{#1}\end{eqnarray}}
\def\eeqan{\end{eqnarray}}

\def\Acknowledgements{\bigskip  \bigskip \begin{center} \begin{large}
             \bf ACKNOWLEDGEMENTS \end{large}\end{center}}

\def\Re{{\cal R \mskip-4mu \lower.1ex \hbox{\it e}\,}}
\def\Im{{\cal I \mskip-5mu \lower.1ex \hbox{\it m}\,}}
\def\nn{\noindent}
\def\ie{{\it i.e.}}
\def\eg{{\it e.g.}}

\def\etal{{\it et al.}}
\def\ibid{{\it ibid}.}
\def\sub#1{_{\lower.25ex\hbox{$\scriptstyle#1$}}}
\def\sul#1{_{\kern-.1em#1}}
\def\sll#1{_{\kern-.2em#1}}  
\def\sbl#1{_{\kern-.1em\lower.25ex\hbox{$\scriptstyle#1$}}}
\def\ssb#1{_{\lower.25ex\hbox{$\scriptscriptstyle#1$}}}
\def\sbb#1{_{\lower.4ex\hbox{$\scriptstyle#1$}}}

\def\to{\rightarrow}
\def\dk{\ifmmode \Delta\kappa\else $\Delta\kappa$\fi}
\def\sigt{\ifmmode \tilde\sigma\else $\tilde\sigma$\fi}
\def\mh{\ifmmode m\sbl H \else $m\sbl H$\fi}
\def\mch{\ifmmode m_{H^\pm} \else $m_{H^\pm}$\fi}
\def\mt{\ifmmode m_t\else $m_t$\fi}
\def\mc{\ifmmode m_c\else $m_c$\fi}
\def\mz{\ifmmode M_Z\else $M_Z$\fi}
\def\mw{\ifmmode M_W\else $M_W$\fi}
\def\mws{\ifmmode M_W^2 \else $M_W^2$\fi}
\def\mhs{\ifmmode m_H^2 \else $m_H^2$\fi}   
\def\mzs{\ifmmode M_Z^2 \else $M_Z^2$\fi}
\def\mts{\ifmmode m_t^2 \else $m_t^2$\fi}
\def\mcs{\ifmmode m_c^2 \else $m_c^2$\fi}
\def\mchs{\ifmmode m_{H^\pm}^2 \else $m_{H^\pm}^2$\fi}
\def\ztwo{\ifmmode Z_2\else $Z_2$\fi}
\def\zone{\ifmmode Z_1\else $Z_1$\fi}
\def\mtwo{\ifmmode M_2\else $M_2$\fi}
\def\mone{\ifmmode M_1\else $M_1$\fi}
\def\tb{\ifmmode \tan\beta \else $\tan\beta$\fi}
\def\xw{\ifmmode x\sub w\else $x\sub w$\fi}
\def\ch{\ifmmode H^\pm \else $H^\pm$\fi}
\def\lum{\ifmmode {\cal L}\else ${\cal L}$\fi}
\def\inpb{\ifmmode {\rm pb}^{-1}\else ${\rm pb}^{-1}$\fi}
\def\infb{\ifmmode {\rm fb}^{-1}\else ${\rm fb}^{-1}$\fi}
\def\epem{\ifmmode e^+e^-\else $e^+e^-$\fi}
\def\ppb{\ifmmode \bar pp\else $\bar pp$\fi}

\def\bsg{\ifmmode b\rightarrow s\gamma \else $b\rightarrow s\gamma$\fi}
 
\newskip\zatskip \zatskip=0pt plus0pt minus0pt
\def\matth{\mathsurround=0pt}

\def\atversim#1#2{\lower0.7ex\vbox{\baselineskip\zatskip\lineskip\zatskip
  \lineskiplimit 0pt\ialign{$\matth#1\hfil##\hfil$\crcr#2\crcr\sim\crcr}}}

\begin{document}
\rightline{\vbox{\halign{&#\hfil\cr
&SLAC-PUB-7154\cr
&May 1996\cr}}}
\vspace{0.8in} 
\Title{Anomalous Chromoelectric and Chromomagnetic Moments of the Top Quark 
at the NLC
}
\bigskip
\Author{Thomas G. Rizzo\doeack}
\Address{\SLAC}
\bigskip
\begin{Abstract}
 
The production of top quark pairs in association with a hard gluon in $e^+e^-$ 
collisions at the NLC provides an opportunity to probe for anomalous 
$t\bar tg$ couplings, \eg, the chromoelectric and chromomagnetic moments of 
the top. We demonstrate that an examination of the energy spectrum of the 
additional gluon jet can yield strong constraints on the size of these two 
anomalous couplings. The results are shown to be quite sensitive to the cut 
on the minimum gluon jet energy needed to remove events where the final 
state $b$-quark radiates strongly. The possibility of using additional 
observables to improve the sensitivity to anomalous couplings is briefly 
discussed.

\end{Abstract}
\bigskip
\vskip1.0in
\begin{center}
To appear in {\it Physics and Technology of the Next Linear 
Collider}, eds. D.\ Burke and M.\ Peskin, reports submitted to Snowmass 1996.
\end{center}
\bigskip

\def\thefootnote{\fnsymbol{footnote}}
\setcounter{footnote}{0}
\newpage
\section{Introduction}

The discovery of the top quark with a mass consistent with the expectations 
from precision electroweak measurements is a major triumph for the Standard 
Model(SM). 
However, due to the fact that it is so heavy, $m_t=175\pm 9$ GeV{\cite {mass}}, 
the top itself has been 
proposed as a window for new physics {\it beyond} the SM. One obvious 
scenario is a modification of  the interactions of the top quark with the 
conventional gauge bosons, \ie, the $W$, $Z$, $\gamma$, and $g$, whose 
couplings to top may differ in detail from those 
anticipated by the SM. This possibility has lead to a substantial effort over 
the past few years{\cite {big}} investigating potential anomalous couplings 
of the top as well as other third generation fermions. 
In the case of the strong interactions of the top, the lowest dimensional 
gauge-invariant 
operator representing new top quark physics and conserving $CP$
that we can introduce is the anomalous chromomagnetic moment, which we can 
parameterize via a dimensionless quantity  
$\kappa$. On the otherhand, the corresponding chromoelectric moment, 
parameterized by $\tilde \kappa$, violates $CP$ and arises from an operator 
of the same dimension. 
In this modified version of QCD for the top quark, the tree-level three-point 
$t\bar tg$ interaction Lagrangian takes the form 
\begin{equation}
{\cal L}=g_s\bar t T_a \left( \gamma_\mu+{i\over {2m_t}}\sigma_{\mu\nu}
(\kappa-i\tilde \kappa \gamma_5)q^\nu\right)t G_a^\mu \,,
\end{equation}
where $g_s$ is the strong coupling constant, $m_t$ is the top quark mass, 
$T_a$ are the color generators, $G_a^\mu$ is the gluon field and 
$q$ is the outgoing gluon momentum. Due to the non-Abelian nature of 
QCD, a four-point, dimension-five $t\bar tgg$ interaction is also generated, 
but this will 
not concern us in the present work since it only contributes to the process of 
interest at higher order. As has been discussed in the literature, if either or 
both of $\kappa,\tilde \kappa$ are 
sufficiently large in magnitude their effects can be probed through top pair 
production processes at both $e^+e^-${\cite {nlc}} and hadron{\cite {had}} 
colliders. The purpose of the present work is to consider the sensitivity 
of $t\bar tg$ production in $e^+e^-$ collisions at the NLC to non-zero 
values of $\kappa$ and $\tilde \kappa$. 

\section{Analysis}

In our analysis we will only consider how the top quark anomalous couplings can 
modify the energy distribution of the extra gluon jet associated with top pair 
production. In principle, other observables may be available that are also 
sensitive to these couplings; these are generally beyond the scope of the 
present work, but one example will be discussed later. 
The basic cross section formulae and analysis procedure can be found 
in Ref.{\cite {nlc}}. Here, we go beyond this initial study in several ways: 
($i$) we generalize the form of the $t\bar tg$ coupling to allow for 
the possibility of a 
sizeable chromoelectric moment, $\tilde \kappa$. The incorporation of 
$\tilde \kappa \neq 0$ into the expressions for the differential cross section 
in Ref.{\cite {nlc}} is rather 
straightforward and can be accomplished by the simple substitution 
$\kappa^2 \to \kappa^2+\tilde \kappa^2$ made universally. 
Note that since a non-zero value of $\tilde \kappa$ 
produces a $CP$-violating interaction it appears only quadratically in the 
expression for the gluon energy distribution since this is a $CP$-conserving 
observable. Thus, in comparison 
to $\kappa$, we anticipate a greatly reduced sensitivity to the value of 
$\tilde \kappa$. ($ii$) We use updated 
expectations for the available integrated luminosities of the NLC at various 
center of mass energies as well as an updated efficiency($\simeq 100\%$) for 
identifying top-quark pair production events. Both of these changes obviously 
leads to a direct increase in statistical power compared to 
Ref.{\cite {nlc}} ($iii$) We lower the cut placed 
on the minimum gluon jet energy, $E_g^{min}$, in performing the energy spectrum 
fits. The reasons 
for having such a cut are two-fold. First, a minimum gluon 
energy is required to identify the event as $t\bar tg$. The cross section 
itself is infra-red singular 
though free of co-linear singularities due to the finite top quark mass. 
Second, since the top decays rather quickly, 
$\Gamma_t\simeq 1.45$ GeV, we need to worry about `contamination' from 
the additional gluon radiation 
off of the $b$-quarks in the final state. Such events can be effectively 
removed from our sample if we require that $E_g^{min}/\Gamma_t>>1$. In our 
past analysis we were overly conservative in our choices for $E_g^{min}$ in 
order to make 
this ratio as large as possible, \ie, we assumed $E_g^{min}=50(200)$ GeV for 
an NLC with a center of mass energy of 500(1000) GeV. It is now believed that 
we can with reasonable justification soften these cuts to at least as low a 
value as 
25(50) GeV for the same center of mass energies{\cite {orr}}, with a potential  
further softening of the cut at the higher energy machine being possible. 
Due to the dramatic 
infra-red behaviour of the cross section, this change in the cuts leads not 
only to an increased statistical power but also to a longer lever arm to 
probe events with very large gluon jet energies which have the most 
sensitivity to the presence of anomalous couplings. Combining all these 
modifications, as one might expect, 
we find constraints which are substantially stronger than what was obtained in 
our previous analysis{\cite {nlc}}. 

\vspace*{-0.5cm}
\nn
\begin{figure}[htbp]
\centerline{
\psfig{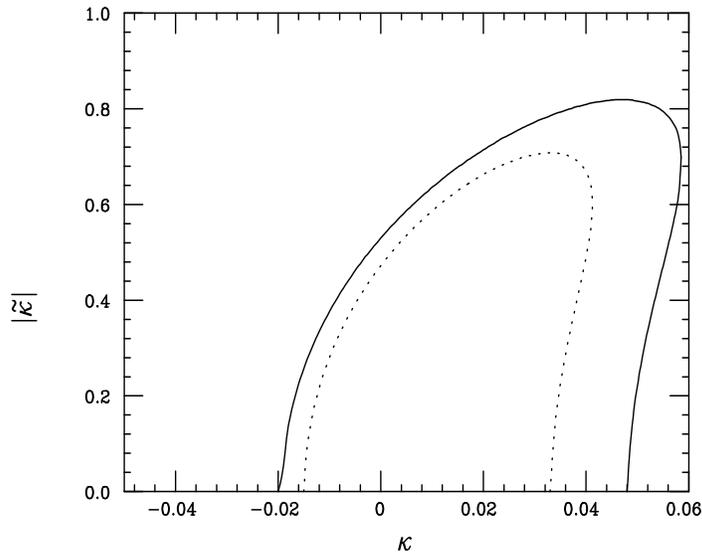}}
\vspace*{-1cm}
\caption{\small $95\%$ CL allowed region in the $\kappa-\tilde \kappa$ plane 
obtained from fitting the gluon spectrum above $E_g^{min}$=25 GeV at a 
500 GeV NLC assuming an integrated luminosity of 50(solid) or 100(dotted) 
$fb^{-1}$.}
\end{figure}
\vspace*{0.4mm}

As in Ref.{\cite {nlc}}, our analysis follows a Monte Carlo approach employing 
statistical errors only. For a given $e^+e^-$ center of mass energy, a binned 
gluon jet spectrum is generated for energies above $E_g^{min}$ assuming that 
the SM is correct. The bin widths are fixed to be $\Delta z=0.05$ where 
$z=2E_g/\sqrt s$ for all values of $\sqrt s$, with the number of bins thus 
determined by the values of the top mass($m_t$=175 GeV), $\sqrt s$ and 
$E_g^{min}$. All calculations are 
performed only in the lowest order. As an example, at a 500 GeV 
NLC with $E_g^{min}$=25 GeV, there are 8 energy bins for the gluon energy 
spectrum beginning at $z=0.10$; 
the last bin covers the range above $z=0.45$. After the Monte Carlo data 
samples are generated, we 
perform a fit to the general expressions for the $\kappa-\tilde \kappa$ 
dependent spectrum and obtain the 
95$\%$ CL allowed region in the $\kappa-|\tilde \kappa|$ plane. (Note that 
only the 
absolute value of $\tilde \kappa$ occurs due to the reasons described above.)

\vspace*{-0.5cm}
\nn
\begin{figure}[htbp]
\centerline{
\psfig{figure=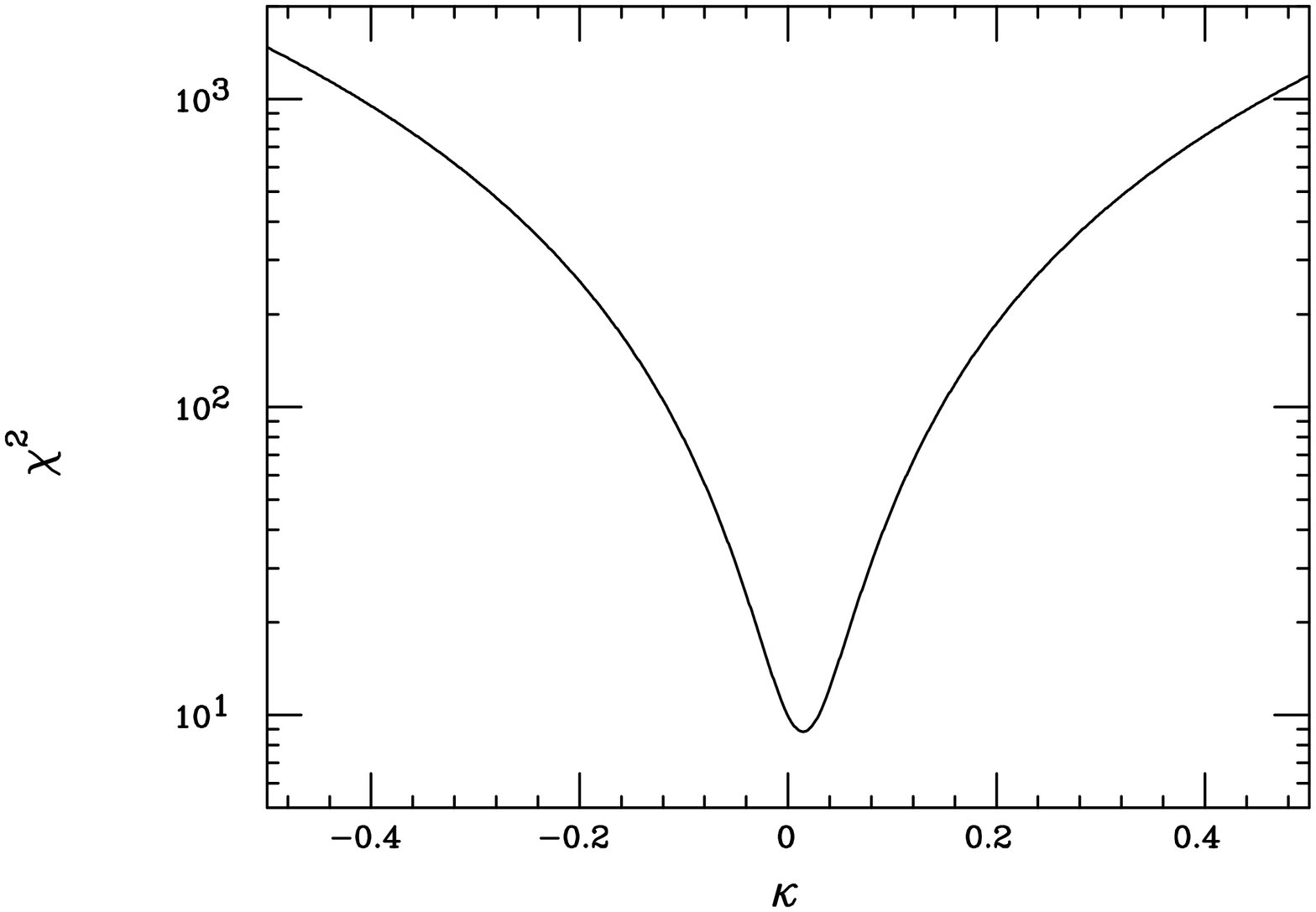,height=9.1cm,width=9.1cm,angle=-90}
\hspace*{-5mm}
\psfig{figure=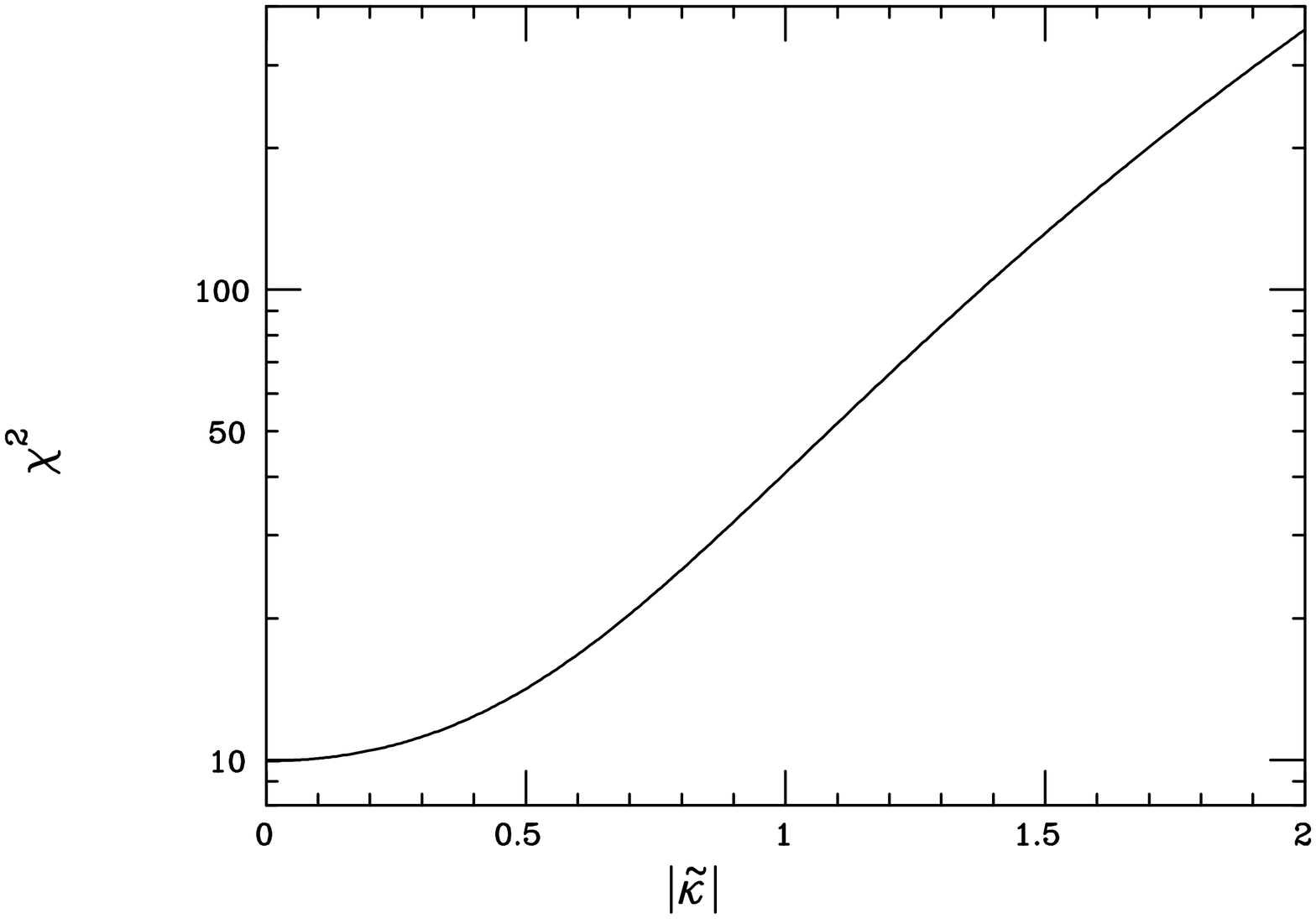,height=9.1cm,width=9.1cm,angle=-90}}
\vspace*{-1cm}
\caption{\small $\chi^2$ plots for the 500 GeV NLC with a luminosity of 50 
$fb^{-1}$ corresponding to Fig. 1. Only one of $\kappa$ or $\tilde \kappa$ is 
assumed to be non-zero at a time.}
\end{figure}
\vspace*{0.4mm}

Fig. 1 shows the results of this procedure for a 500 GeV NLC with a cut of 
$E_g^{min}$=25 GeV for two different integrated luminosities. As expected, 
excellent constraints on $\kappa$ are now obtained but those on 
$\tilde \kappa$ are more than an order of magnitude weaker. A doubling of the 
integrated luminosity from 50 to 100 $fb^{-1}$ decreases the size of the 
allowed region by about 40$\%$. We note that in the previous study 
only extremely 
poor constraints on $\kappa$ were obtained at a 500 GeV $e^+e^-$ collider, 
$-1.98\leq \kappa \leq 0.44$, due to the presence of a degenerate minima in 
the $\chi^2$ distribution. Now, with the both the increased luminosity and 
top-tagging efficiencies, as well as the longer lever arm in energy, these 
previous difficulties are circumvented.  This is explicitly shown by the 
$\chi^2$ plots in Fig. 2. In the case of a non-zero $\kappa$ a very sharp 
minimum is found whereas the rise in $\chi^2$ is slow when $\tilde \kappa$ is 
non-zero. In neither case is there any evidence for a second minimum. This 
lack of degeneracy is found to hold for all the other cases we have 
considered.

\vspace*{-0.5cm}
\nn
\begin{figure}[htbp]
\centerline{
\psfig{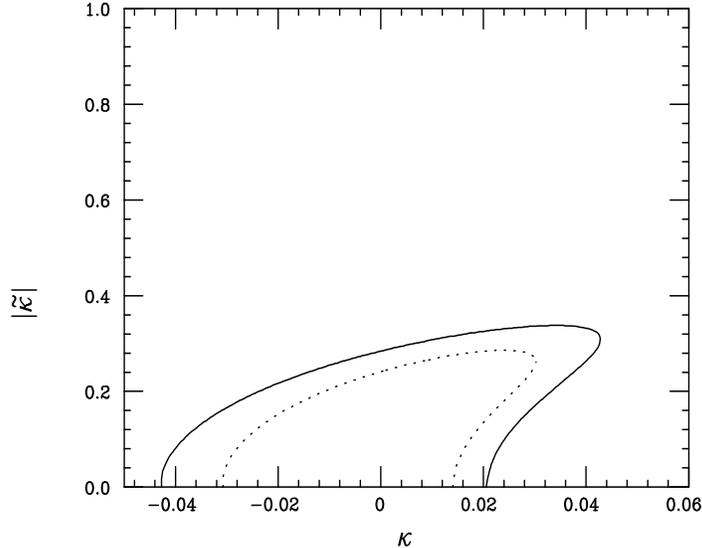}}
\vspace*{-1cm}
\caption{\small Same as Fig. 1, but for a 1 TeV collider with $E_g^{min}$=50 
GeV and luminosities of 100(solid) and 200(dotted) $fb^{-1}$. Note that the 
allowed region has been significantly compressed downward in comparison to 
Fig. 1.}
\end{figure}
\vspace*{0.4mm}

Going to a higher energy leads to several simultaneous effects. First, since 
the cross section 
approximately scales like $\sim 1/s$ apart from phase space factors, 
a simple doubling of the 
collider energy induces a reduction in statistics unless higher integrated 
luminosities are available to compensate. Second, the sensitivity 
to the presence of 
non-zero anomalous couplings is enhanced at higher energies, roughly scaling 
like $\sim \sqrt s$ for $\kappa$ and, correspondingly, like $\sim s$ for 
$\tilde \kappa$ {\it assuming} the same available statistics at all energies. 
In Fig. 3 we show the results of our analysis at a 1 TeV NLC 
for $E_g^{min}$=50 GeV; the corresponding case where we maintain the jet 
energy cut of  $E_g^{min}$=25 GeV is shown in Fig. 4. (Note that in our 
previous analysis, we obtained the $95\%$ CL bound $-0.12\leq \kappa \leq 0.21$ 
for this center of mass energy and an integrated luminosity of 200 
$fb^{-1}$.) For $E_g^{min}$=50(25) 
GeV, the energy range is divided into 15(16) $\Delta z=0.2$ bins beginning at 
$z=0.10(0.05)$ with the last bin covering the range $z\geq 0.80$. We see from 
these figures that by going to higher energy we drastically compress the 
allowed range of $\tilde \kappa$ while the improvement for $\kappa$ is not as 
great. Lowering the energy cut is seen to lead to a far greater reduction in 
the size of the 95$\%$ CL allowed region than is a simple doubling 
of the integrated 
luminosity. This demonstrates that if improvements to the present analysis 
are to be made it is very important to explore just how 
low the $E_g^{min}$ can be placed and still remove the contamination from gluon 
radiation off of final state $b$-quarks. 
Comparing the 200 $fb^{-1}$ result in Fig. 4 with the 
corresponding case of 50 $fb^{-1}$ in Fig. 1, we see that our approximate 
scaling laws are observed to hold rather well. 

\vspace*{-0.5cm}
\nn
\begin{figure}[htbp]
\centerline{
\psfig{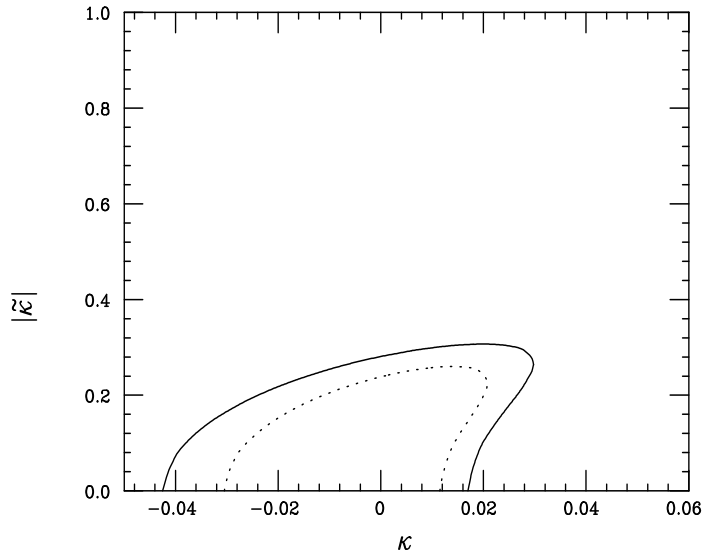}}
\vspace*{-1cm}
\caption{\small Same as Fig. 3 but with $E_g^{min}=25$ GeV.}
\end{figure}
\vspace*{0.4mm}

As a final example, we consider a $\sqrt s$=1.5 TeV NLC with 
$E_g^{min}$=75 GeV. This requires 16 $\Delta z=0.05$ bins beginning at 
$z=0.10$ to cover the entire gluon jet energy spectrum; the final bin 
covers the range $z\geq 0.85$. The results of this analysis are shown in 
Fig. 5 assuming integrated luminosities of 200 and 300 $fb^{-1}$. The $95\%$ 
CL allowed region is seen to be further compressed in the vertical 
direction, \ie, to smaller 
values of $\tilde \kappa$, than that found in the 1 TeV NLC case with either 
$E_g^{min}$=25 or 50 GeV. However, the overall size of the allowed region for 
the 1.5 TeV collider with $E_g^{min}=$75 GeV is comparable to that of the 1 
TeV NLC with $E_g^{min}=$25 GeV. Softening the $E_g^{min}$ cut at the 1.5 TeV 
collider will yield even stronger constraints.

\vspace*{-0.5cm}
\nn
\begin{figure}[htbp]
\centerline{
\psfig{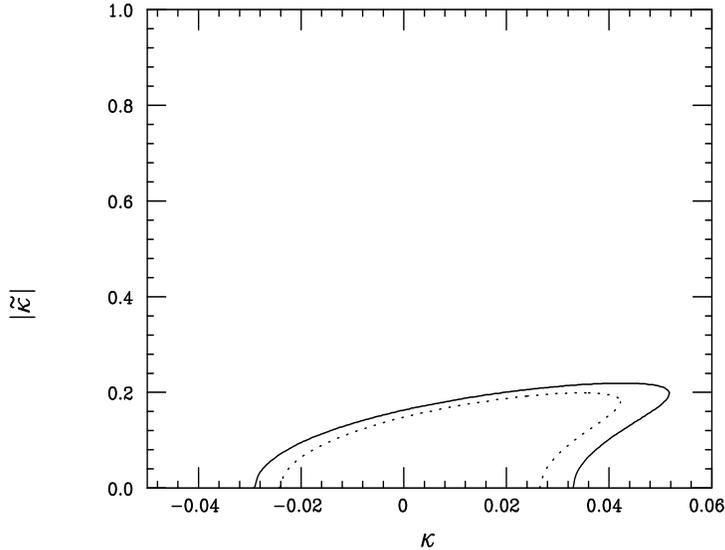}}
\vspace*{-1cm}
\caption{\small Same as Fig. 1 but for a 1.5 TeV NLC with $E_g^{min}$=75 GeV 
for luminosities of 200(solid) and 300(dotted) $fb^{-1}$.}
\end{figure}
\vspace*{0.4mm}

\section{Discussion and Conclusions}

We have presented an updated analysis of the constraints imposed on the 
chromoelectric and chromomagnetic moments of the top quark by detailed 
measurements of the gluon jet energy spectrum associated with the process 
$e^+e^- \to t\bar tg$ at the NLC for various center of mass energies. The 
value of the cut on the gluon energy was shown to play a key role in obtaining 
strong bounds on these anomalous couplings. These results may be strengthened 
in the future if we find that the $E_g^{min}$ cut can be further softened. 

If we are to improve the sensitivity to anomalous couplings we must introduce 
additional observables. The large initial electron beam polarization, 
$P=90\%$, may play an important role in this regard. To get an idea of how 
this might work, we can 
construct an asymmetry using the Monte Carlo data generated above by asking 
how the number of events in a fixed energy bin is altered when one changes from 
left-handed to right-handed polarization, \ie, 
$A(z)=[N_L(z)-N_R(z)]/[N_L(z)+N_R(z)]$. For a $\sqrt s$=500 GeV NLC, $A$ is 
found to be approximately $z$ independent with a value near $39\%$ in the SM 
and will in 
general be $\kappa, \tilde \kappa$ dependent if anomalous couplings are 
present. We thus can repeat our analysis for the 500 GeV case now including 
the values of $A$ in the fit. This is shown in Fig. 6. Unfortunately including 
$A$ has not drastically reduced the size of the $95\%$ CL allowed region as we 
might have hoped since $A$ is only rather weakly dependent on the anomalous 
couplings. Thus we must seek out other observables, particularly those 
which are sensitive to the $CP$-violating chromoelectric moment interactions. 
This analysis is currently underway.

\vspace*{-0.5cm}
\nn
\begin{figure}[htbp]
\centerline{
\psfig{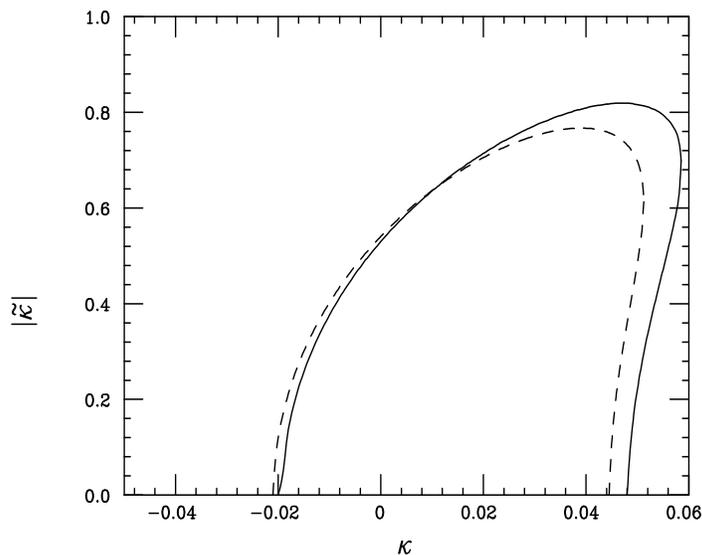}}
\vspace*{-1cm}
\caption{\small The solid curve is the same as in Fig. 1 for a 500 GeV NLC 
with $E_g^{min}$=25 GeV and a luminosities of 50 $fb^{-1}$. The dashed curve 
included the polarization asymmetry into the fit assuming an initial beam 
polarization of $90\%$.}
\end{figure}
\vspace*{0.4mm}

\Acknowledgements

The author would like to thank P. Burrows, D. Atwood, A. Kagan and 
J.L. Hewett for discussions related to this work.

%
\def\MPL #1 #2 #3 {Mod.~Phys.~Lett.~{\bf#1},\ #2 (#3)}
\def\NPB #1 #2 #3 {Nucl.~Phys.~{\bf#1},\ #2 (#3)}
\def\PLB #1 #2 #3 {Phys.~Lett.~{\bf#1},\ #2 (#3)}
\def\PR #1 #2 #3 {Phys.~Rep.~{\bf#1},\ #2 (#3)}
\def\PRD #1 #2 #3 {Phys.~Rev.~{\bf#1},\ #2 (#3)}
\def\PRL #1 #2 #3 {Phys.~Rev.~Lett.~{\bf#1},\ #2 (#3)}
\def\RMP #1 #2 #3 {Rev.~Mod.~Phys.~{\bf#1},\ #2 (#3)}
\def\ZP #1 #2 #3 {Z.~Phys.~{\bf#1},\ #2 (#3)}
\def\IJMP #1 #2 #3 {Int.~J.~Mod.~Phys.~{\bf#1},\ #2 (#3)}

\end{document}